# Evidence for a disaggregation of the universe

Antonio Alfonso-Faus

Departamento de Aerotécnia
Madrid Technical University (UPM), Spain
April, 2011. E-mail: aalfonsofaus@yahoo.es

**Abstract:** Combining the kinematical definitions of the two dimensionless parameters, the deceleration $q(x)$ and the Hubble $t_0 H(x)$, we get a differential equation (where $x = t/t_0$ is the age of the universe relative to its present value $t_0$). First integration gives the function $H(x)$. The present values of the Hubble parameter $H(1)$ [approximately $t_0 H(1) \approx 1$], and the deceleration parameter [approximately $q(1) \approx -0.5$], determine the function $H(x)$. A second integration gives the cosmological scale factor $a(x)$. Differentiation of $a(x)$ gives the speed of expansion of the universe. The evolution of the universe that results from our approach is: an initial extremely fast exponential expansion (inflation), followed by an almost linear expansion (first decelerated, and later accelerated). For the future, at approximately $t \approx 3t_0$ there is a final exponential expansion, a second inflation that produces a disaggregation of the universe to infinity. We find the necessary and sufficient conditions for this disaggregation to occur. The precise value of the final age is given only with one parameter: the present value of the deceleration parameter $q(1) \approx -0.5$]. This emerging picture of the history of the universe represents an important challenge, an opportunity for the immediate research on the Universe. These conclusions have been elaborated without the use of any particular cosmological model of the universe.

# 1. – Introduction

During the year 2002 McInnes [1] advanced the idea that a destruction of the universe in a finite time, by excessive expansion, was possible. A few months later Caldwell [2], following his ideas on phantom energy, went on arriving at the conclusion that something more than the usual vacuum energy, this energy identified by the cosmological constant $\Lambda$, ought to be present too. Then there could be enough extra pressure, more than thought, to expand the universe to infinity, in a finite time. The next year 2003, Caldwell et al. [3] arrived at the conclusion that, with only 50% more expanding pressure than the one given by the usual $\Lambda$, the so called phantom energy would cause a cosmic "doomsday" at about 2.66 times the present age of the universe. This year 2011 Li and Wu [4], examining the cosmic acceleration with the latest Union2 supernova data, found that the present acceleration of the expansion of the universe is increasing. Therefore disaggregation was not ruled out by observation. Finally, also this year 2011, Akarsu and Dereli [5] presented cosmological models with linearly varying deceleration parameter $q(x)$. In this case the conclusion was also that the universe ends with a big-rip, as presented by the previous authors.

About the same time, this year 2011 we have presented [6] a similar method to use the measured values of the deceleration parameter $q$, and arrived at the history of the size of the universe, the speed of expansion etc. The assumed law for $q(x)$ used in [6] is different from the one used by the previous authors [5], but the same result of disaggregation of the universe in a finite time was found.

In the present work we refine the approach we took in [6] in order to get the necessary and sufficient conditions to have a disaggregation of the universe, in a finite time. In this way we also get initial inflation, as proposed by Guth [7] and Linde [8]. They introduced the idea of an initial exponential expansion during a very short period of time, an initial inflation related to the vacuum properties (may be of space-time). This evidently helps the initial big-bang model of the universe to increase its size, followed by the "normal" expansion. Today the inflationary hypothesis is getting more and more support due to the agreements seen with its predictions: flat universe, critical density, properties of the cosmic background radiation and so on. Usually this inflation epoch is maintained to have occurred after the "big-bang". We have not seen here any need to day for the survival of the big-bang idea: inflation from an initial quantum black hole (Planck´s type) is enough to do the job. I refer to the fact that the Planck´s quantum black hole physical properties (mass, length, time and so on), when multiplied by the dimensionless scale factor $10^{61}$, give the present physical properties of our universe! [9].

The point is that, in order to break the fast initial exponential expansion, inflation, gravitational attraction has to enter into the picture. Something like artificial fireworks that explode in an accelerated expansion, followed by more expansion but decelerated. This picture would give us for today a decelerated expansion for the universe. But this is not what is observed: today we see an accelerated expansion, which is occurring back in time from about half the present age of the universe [10]. If $a(t)$ is the cosmological scale factor, the "deceleration" parameter $q$ has been defined as

$$q(t) = - \ddot{a}(t)\, a(t) / \dot{a}(t)^2 \qquad (1)$$

One would expect that, after inflation, the universe was left still expanding but slowly decelerating. Then the definition of $q$ in (1) would give (with $\ddot{a}(t) < 0$) a value of $q > 0$.

The reason to include a negative sign in (1) was to get a positive value for $q$, even to call $q$ the deceleration parameter, a positive dimensionless number. But nature keeps giving us surprises. About ten years ago it was found that the universe is in a state of *accelerated expansion,* and we know today that this state of acceleration started back in time at about one half the age of the universe [10]. Initially we needed the cosmological constant Λ to balance gravitation, to have a push to explain the expansion of the universe. This pressure obviously implies energy, and the constant Λ has been linked with this energy. In many instances it has been identified with the vacuum energy. But now we see that there is even more energy increasing the acceleration of the expansion of the universe.

## 2. – The Magic equation

To have a measure of the expansion rate of the universe, a well known parameter H (the Hubble parameter) has been defined as

$$H(t) = \dot{a}(t)/a(t) = d[\ln a(t)]/dt \qquad (2)$$

The combination of the definitions (1) and (2) give the important cosmological equation

$$\dot{H} + [1 + q(t)]H^2 = 0 \qquad (3)$$

This equation is equivalent to the following one

$$d(1/H)/dt = [1 + q(t)] \qquad (4)$$

No particular cosmological model of the universe is necessary to integrate this equation. Only the function $q(t)$. We call it magic because it contains all the geometry and kinematics of the universe, as we will see. On the other hand, enough measurements of $q$ are already known today for a reasonable, and plausible, integration of (3).

The dimensionless parameter $x$ is defined as $t/t_0$, the age of the universe relative to the present age ($x = 1$). Integrating (3) between the limits $x = 0$ and $x = 1$, and taking into account that the present value ($x = 1$) of $H(1)t_0$ is about 1 we get the result

$$0 \approx \int q(x)\,dx \qquad (5)$$

where the integration is from the origin (x = 0) to the present age (x = 1). Here we have introduced an assumption that is very well substantiated, and with a later check. The function $H(x)t_0$ is about 1 today, as measured. It is roughly inversely proportion to the age of the universe. Therefore close to the origin ( $x \approx 0$ ) its value is very high, $H(0)t_0 >> H(1)t_0$. And its inverse must be very close to zero. Then the integral of (4) between 0 and 1 is inevitably the condition (5). Back in time the expected value for $q_1 = q(0)$ after inflation, and very near the initial stages, should be found as a limiting process [11] and is close to 0.5 (at a very high red shift). The present value of $q$ is roughly about $q_0 \approx - 0.5$. In between this two ages (the initial $t \approx 0$ and today $t_0 \approx 1.37 \; 10^{10}$ years) one must have a value of zero for $q$. This is because one must cross from a positive value (deceleration after inflation) to a negative one for today (acceleration). A zero acceleration $q_{acc} = 0$ at about $x \approx 0.5$ has been found [10] followed by an acceleration up to the present time ($q_0 \approx - 0.5$). Here we will prove that the condition (5) inevitably predicts a zero value for $q$ at the central point in age x = 0.5. We then have three values for $q$: $q_1 \approx 0.5$, $q_{acc} (0.5)= 0$, and $q_0 \approx -0.5$, corresponding to the ages $x = 0, 0.5, and 1$. The condition (5) is fulfilled by any function of $q(x)$, between x = 0 and x = 1, provided that its integration between 0 and 0.5 be cancelled by its integration between 0.5 and 1. For example, a linear one does the job, which is the choice in [5]. The condition (5) has been deduced here as being a necessary condition for $q(x)$, if and only if, the present value of the Hubble parameter is $H(1) \approx 1/t_0$, and we do know that this is precisely the case. It is then an experimental necessary condition for $q(x)$ to hold between the initial stage (x=0) of the universe and today (x=1). No particular model of the universe has been used here to arrive at this condition.

We see now that the integration of (4) is given by

$$1/[(t_0 H(x)] \approx 1 + \int [1+q(x)] \; dx \qquad (6)$$

where the limits of integration are from x = 1 to a general x.

With the definition for H in (2) and using the dimensionless $x$ the equation (6) transforms to

$$d\,[\ln a(x)]/dx = \left(1 + \int [1+q(x)]\,dx\right)^{-1} \qquad (7)$$

with the limits of integration between 1 and x. Now we need the function $q(x)$ to integrate (7). There is no need to know it between x = 0 and x = 1, as long as the condition (5) is satisfied. We choose now a linear variation of $q(x)$ between x = 1 ($q = q_0$) and a general x with $q(x)$. This straight line has an inclination α defined by m = tg α. Then we get for $q(x)$

$$q(x) = m(x - 1) + q_0 \qquad (8)$$

Applying the condition (5) with the expression (8) for $q(x)$ we get

$$0 \approx m(x^2/2 - x) + q_0 x \qquad (9)$$

$$\text{i.e.} \quad 0 \approx m(x/2 - 1) + q_0 \qquad (10)$$

where we see that the initial stages, x ≈ 0, satisfy this condition (9), as they should. Given that today, x = 1, this condition must be satisfied too, we get the important relation between m and $q_0$

$$m \approx 2\,q_0 \qquad (11)$$

This condition implies that the function $q(x)$ defined in (8) must be

$$q(x) \approx q_0\,(2x - 1) \qquad (12)$$

It is a very important result. It predicts two particular values of the parameter q. Independently of the precise value for today, $q_0$, it gives us

$$x = 1/2, \quad q(1/2) \approx 0$$

$$x = 0, \quad q(0) \approx -\,q_0 \qquad (13)$$

These two predictions are very well satisfied by observations, [5] and [11]. They imply symmetry in the function $q(x)$ such that our present age, where the earth is populated by the human race, is twice the age at which the acceleration of the expansion of the universe was zero. Also, independently of the actual value of $q_0$ for today, the initial value $q(0)$ is $-q_0$. The following Fig. 1 is the plot of the function $q(x)$, with the choice $q_0 = -.5$.

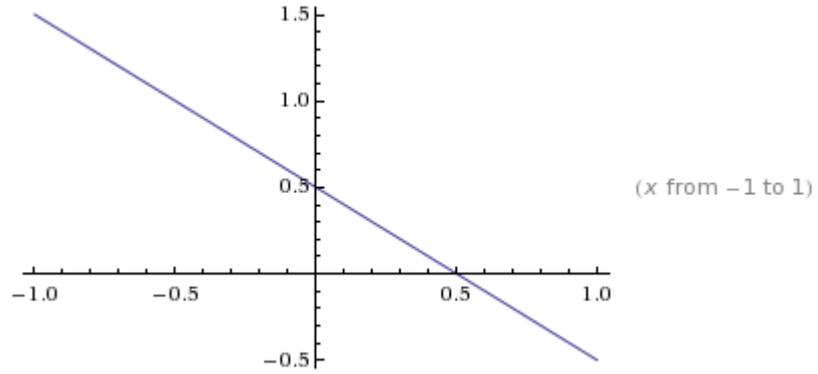

Fig. 1 Plot of $q(x) = .5 - x$ (for the choice $q_0 = -.5$).

Integrating (7) one gets

$$\ln a(x) = \int (1 + \int [1+q(x)]\, dx)^{-1} dx \qquad (14)$$

where the two integrals have the same limits between 1 and x. From here we get

$$a(x)/a(1) = \exp \int (1 + \int [1+q(x)]\, dx)^{-1} dx \qquad (15)$$

We have now the opportunity to find the vertical asymptotes for the plot of $a(x)$ in terms of $x$, i.e., the history (past and future) of the cosmological scale factor of the universe. We get them by imposing the condition

$$1 + \int [1+q(x)]\, dx = 0 \qquad (16)$$

Then, using (12) we integrate (16) to obtain

$$x + q_0 (x^2 - x) = 0 \qquad (17)$$

We have discovered two vertical asymptotes for $a(x)$. They occur at the time $x_d$, one in the past $(x = 0$, that we identify with inflation), and the other one in the future

$$x_d = 1 - 1/q_0 \qquad (18)$$

(Here the sub index d stands for "disaggregation", or "doomsday"). The present observations give for $q_0$ a negative value (acceleration) and it is around $-0.5$. Then $x_d$ is always $> 1$, always in the future, and it is around 3. The end of the universe is here foreseen to happen at about $t_d \approx 4 \; 10^{10}$ years. The conclusion is that the first inflation at $x=0$ implies the birth of the universe, and the second one at $x_d = 3$ implies its death.

The exponential expansion introduced by Guth [7] and Linde [8], inflation, can be interpreted as a very short period of time such that the Hubble parameter H is constant with time. And this must have occurred very close to the initial stages of the universe. From (12) we see that this condition implies for $q$ the initial value $q = -q_0 \approx 0.5$. Given that this must have occurred during a very short period of time, as is the case for inflation, the expected value for $q$ after inflation, and very near the initial stages, should be found as a limiting process [11] and it is in fact close to 0.5. In the Fig. 2 of the next section we have the plot of $a(t)/a_0$ in the vertical axis versus the parameter $x = t/t_0$ in the horizontal axis.

## 3. – The graphical solutions

The integration of the equation (15), using the $q(x)$ function in (12) and with the present value for $q_0 = -0.5$, gives the following solution:

$$a(x)/a(1) = [2x/(3-x)]^{2/3} \qquad (19)$$

The following fig. 2 gives the graphical plot for this cosmological scale factor *a(x)*:

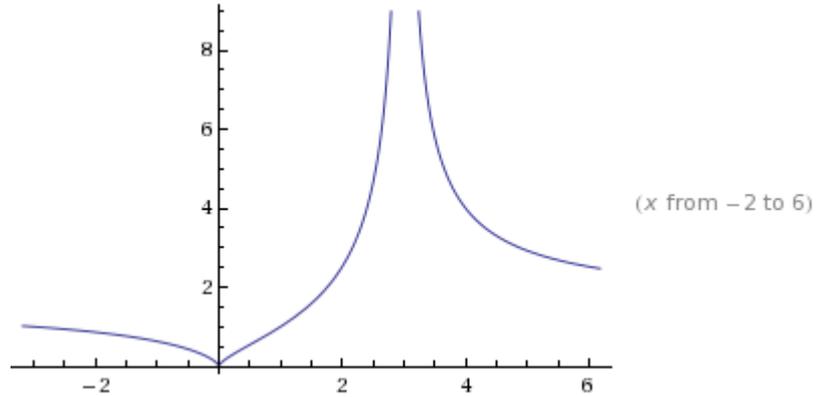

Fig.2 Cosmological scale factor *a(t)/a(1)* versus time $x = t/t_0$

This plot of *a(t)* has a lot of information in it. First we see an almost linear expansion from x = 0 to a little more than x = 1 (today). In the next sections we give an interpretation to the pre big-bang results [12] and [13].

## 4. - The end of the universe

Looking at Fig. 2 for $x = 3$ we see that the universe will spread to infinity at that age. This is an expected extrapolation from the accelerated expansion seen today that goes on until $a(t) \to \infty$. It gives a total lifetime for our universe $t_f \approx 3\ t_0 \approx 4\ 10^{10}$ years. Today we are at about 1/3 of the lifetime. In the following Fig.3 we have the plot of the speed of expansion versus time, as given by the expression

$$a'(x)/a(1) = 4/[2x\ (3-x)^5]^{1/3} \qquad (20)$$

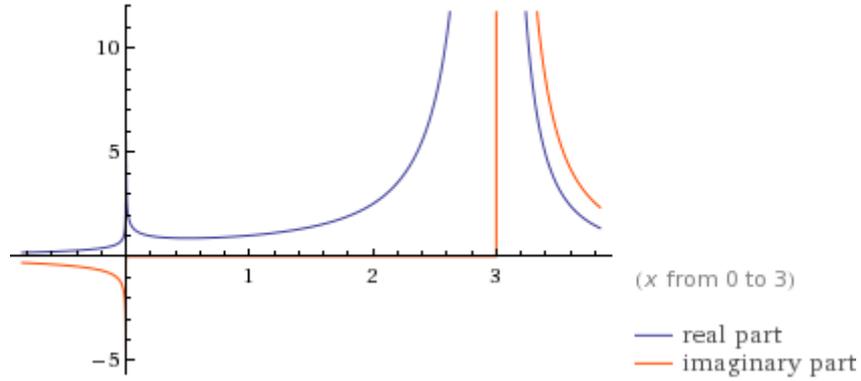

Fig. 3 Vertical axis: speed of expansion relative to the speed of light c. Horizontal axis: time t relative to the present age, $x = t/t_0$.

We see in Fig. 3 that the speed of expansion, relative to the speed of light, is about 1 from the near beginning to a little more at x = 1 (today). From then on it grows up to an infinite value at the end of the life of our universe, for x = 3. But initially (x =0) we have the proposed inflation at a very large speed. The H(x) parameter expression is

$$H(x) = 2/[x(3-x)] \qquad (21)$$

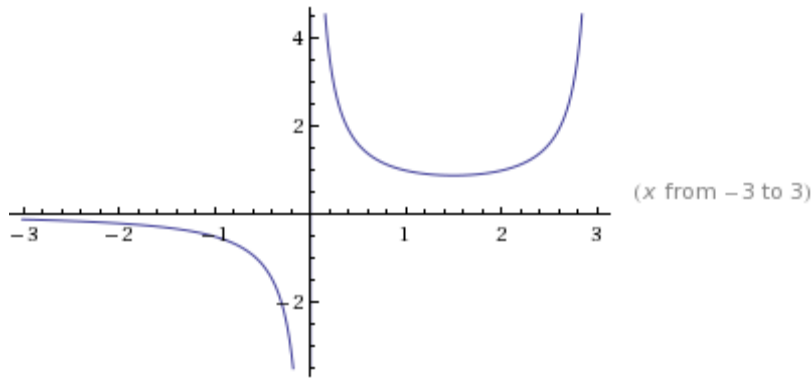

Fig. 4 Vertical axis: dimensionless Hubble parameter $H(x)t_0$ versus age, $x = t/t_0$, at the horizontal axis. There are two infinites for H: one at x = 0, the initial inflation, and the last one at the end of the universe, x = 3.

In Fig 4 we see the evolution of the Hubble parameter H(x) with time. The initial inflation is evident, clearly giving the exponential expansion. The final "inflation" is also evident, showing the disaggregation of everything in our universe. This is an inevitable conclusion. At x = 3 the infinite value for the cosmological scale value in Fig.2, the infinite value for the speed of expansion of the universe in Fig. 3 and the infinite value for the Hubble parameter H in Fig. 4 show the total disaggregation in our universe. If we follow Mach´s ideas on the inertial and gravitational masses, the interaction of any mass m with the rest of the universe is clearly zero. Then there is no mass m and therefore the disaggregation is total and at all levels. It is the end of our universe as we know it.

Finally, fig. 5 gives the acceleration *a´´(x)* that has the expression

$$a´´(x)/a(1) = 2^{8/3}(x - .5) \left[ 1/[x(3-x)^2] \right]^{4/3} \qquad (22)$$

And the corresponding graphical plot

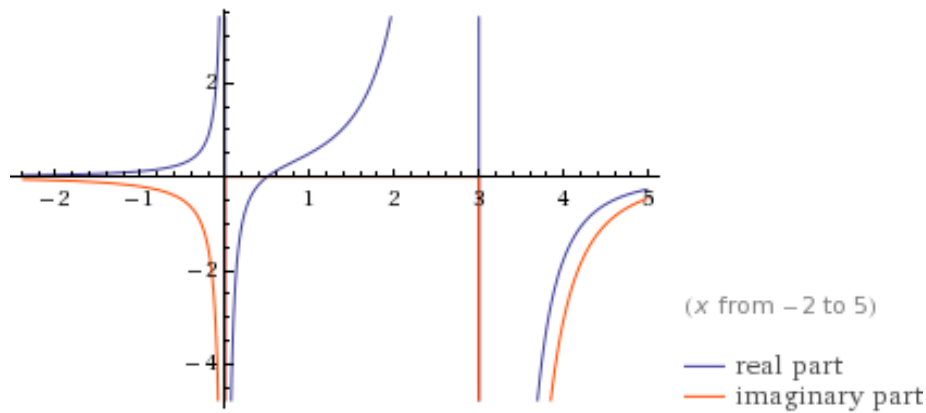

Fig. 5 The acceleration of the expansion of the universe, *a´´(x)*, versus time. The two vertical asymptotes at x = 0 and x = 3 are evident.

## 5. - Black hole formation before x = 0.

Looking at the Figures 1, 2, 3, 4 and 5 it is clear that at an age in the past, x < 0, there is structure in the graphs suggesting the possibility of black hole formation by gravitational collapse, among other possibilities. There are some theoretical and experimental findings in this direction in the literature [12] and [13]. Here we just point out that research in this field may have sense.

## 7. – Conclusions

We have proved here that one needs only one cosmological parameter to find the cosmological scale factor, the size of the universe in terms of time, and its kinematics. It is the deceleration parameter $q$, (and the present value of the Hubble parameter H(1)). We obtain the complete history of the universe, initial inflation, deceleration, acceleration and final expansion to infinity (a second "inflation" at three times the present age of the universe. This has been achieved without the use of the Einstein cosmological equations i.e. the two equations based on his field equations of general relativity.

## 8. – Acknowledgement

I am thanking the owners of the Wolfram Mathematica Online Integrator that I have used to obtain the Figures 1, 2, 3, 4 and 5 of this work.

## 9. – References